\documentclass{article}
\usepackage{graphicx}
\usepackage{amsmath}
\usepackage{amsfonts}
\usepackage{amssymb}

\begin{document}

\title{Current-induced magnetization changes in a spin valve\\due to incoherent emission of non-equilibrium magnons}
\author{V.I. Kozub\\\textit{A.F. Ioffe Physico-Technical Institute}\\\textit{St.-Petersburg 194021, Russian Federation}
\and J. Caro \\\textit{Kavli Institute of Nanoscience Delft}\\\textit{Delft University of
Technology}\\\textit{Lorentzweg 1, 2628 CJ Delft, The
Netherlands}}
\date{\today}
\maketitle
\begin{abstract}
We describe spin transfer in a ferromagnet/normal
metal/ferromagnet spin-valve point contact. Spin is transferred
from the spin-polarized device current to the magnetization of the
free layer by the mechanism of incoherent magnon emission by
electrons. Our approach is based on the rate equation for the
magnon occupation, using Fermi's golden rule for magnon emission
and absorption and the non-equilibrium electron distribution for a
biased spin valve. The magnon emission reduces the magnetization
of the free layer. For anti-parallel alignment of the
magnetizations of the layers and at a critical bias a magnon
avalanche occurs, characterized by a diverging effective magnon
temperature. This critical behavior
can result in
magnetization reversal and consequently to suppression of magnon
emission. However the magnon-magnon scattering can lead to saturation
of magnon concentration at large but finite value. The further behavior
depends on the parameters of the system. In particular, gradual
evolution of magnon concentration followed by the magnetization reversal
is possible. Another scenario corresponds to step-like increase of magnon concentration followed by slow decrease. In the latter case the spike
in differential resistance is expected due to contribution of
electron-magnon scattering. A comparison of the obtained results to existing experimental data and theoretical approches is given.

 \pagebreak
\end{abstract}

\section{Introduction}
The giant magnetoresistance\cite{Baibich} (GMR) of magnetic multilayers is a
strong drop of the electric resistance of such a multilayer on application of
an external magnetic field. This effect arises from the different scattering
strength experienced by the two spin channels making up the spin-polarized
sense current through the multilayer and from a change from antiparallel
alignment of the magnetization of the ferromagnetic layers to parallel
alignment. For antiparallel alignment appreciable scattering of electrons
occurs in both spin channels, for each spin type in those layers where that
type is a minority-spin\cite{spin channels}. For parallel alignment the
majority-spin channel experiences negligible scattering, so that it is a low
resistance channel that short-circuits the high resistance minority-spin channel.

The dual effect of the GMR is a change of the magnetization state of a
magnetic layered structure induced by a spin-polarized bias current traversing
the layers. The mechanism of such a change is transfer of electron spin from
the current to the magnetization of the layers, as proposed by Slonczewski for
a ferromagnet/normal metal/ferromagnet (FM/NM/FM) spin valve\cite{Slonc1}.
This is a trilayer structure, with a NM layer sandwiched between two FM
layers. The spin transfer is predicted\cite{Slonc1} to lead to steady,
coherent precession of the magnetization of the layers and, in the presence of
a uniaxial anisotropy, to switching of the magnetization. These effects should
occur for high current densities ($j=10^{6}-10^{7}$ A/cm$^{\text{2}}$) and
small lateral dimensions (diameter=100-1000 nm). Recently, Katine \textit{et
al}.\cite{Katine} have reported magnetization switching in a spin valve
operating in this regime, which they interpret in the framework of
Slonczewski's theory. However while at small external magnetic fields the behavior
was hysteretic – just in accordance to Slonszewskii theory – at higher
magnetic fields the deviations from Slonszewskii predictions were observed.
Some later, the extended studies of current-driven magnetic evolution were made
by Urazdin et al.\cite{Urazdin}; the authors reported the hysteretic switching at
low external magnetic fields and the reversible behavior in high fields.
It is also important to note that while Slonszewski's theory predict
switching-like evolution of magnetization, our studies of a biased nanometer-scale
Co/Cu/Co spin-valve point contact\cite{Theeuwen} demonstrated
a possibility of gradual evolution; similar features were observed by Katine et al.\cite{Katine}

The approach of Ref. \cite{Slonc1} is semi-classical: the electron
spin is treated quantum mechanically, but spin transfer is derived
from the classical law of angular momentum conservation. Further,
in Ref. \cite{Slonc1} it is ignored that in the spin-transfer
regime the electron system is strongly out of equilibrium.
Actually, for the electron system the spin-transfer regime in a
spin valve is very similar to the regime of point-contact
spectroscopy (PCS) of the electron-magnon interaction in
ferromagnetic metals. PCS of this interaction was theoretically
developed by Kulik and Shekhter \cite{Kulik and Shekhter} for
homogeneous ferromagnetic point contacts, which are short and
narrow constrictions between a pair of three-dimensional
ferromagnetic electrodes. The idea is that in a biased point
contact a bias-dependent non-equilibrium electron distribution is
created. This distribution enables energy relaxation of the
electrons by incoherent emission of magnons, the elementary
excitations of magnetization. Magnon excitation can be detected in
the second derivative of the current-voltage characteristic of the
contact\cite{Akimenko}, which is proportional to the magnon
density of states. In the same spirit, a non-equilibrium electron
distribution generated in a current-biased spin valve should lead
to incoherent emission of magnons in the magnetic layers. To some
extent Berger \cite{Berger} discusses the effect of a
non-equilibrium distribution, but his intuitive approach lacks a
general quantum mechanical basis.

Further theoretical studies of the non-equilibrium dynamics of
magnetic multilayers were presented in Refs.\cite{Bauer} where in
particular the steps to generalize approaches by Slonszewskii and
Berger were made. However these studies were also based on
semiclassical considerations similar to the ones exploited in
\cite{Slonc1}.

In this article we present a consistent quantum mechanical description of spin
transfer in a spin-valve point contact, based on Fermi's golden rule and
taking into account the non-equilibrium electron distribution. Inspired by the
PCS results\cite{Kulik and Shekhter},\cite{Akimenko} we consider emission and
absorption of magnons. However, contrary to PCS, which is concerned with the
direct effect of electron-magnon processes on the electrical transport, we
here focus on the effect of these processes on the magnetization, which can be
strongly reduced by a non-equilibrium population of emitted magnons. A change
of the magnetization due to electron-magnon processes can be probed with the
GMR of the contact, as we recently demonstrated for a biased nanometer-scale
Co/Cu/Co spin-valve point contact\cite{Theeuwen}. We will show that different scenarios of the magnetization evolution are possible including "switching", gradual evolution of magnetization, hysteretic and reversible behavior.

\section{Incoherent emission of magnons}
We consider the spin-valve point contact
This device comprises two metallic electrodes, which are electrically
connected through a small circular orifice (diameter 10-50 nm) in a thin
insulating layer. Adjacent to the insulator, embedded in the left electrode,
there is a spin valve with structure FM($t_{a}\approx2$ nm)/NM($t_{sp,1}%
\approx t_{a}$)/FM($t_{p}\approx100$ nm), the layer thicknesses
being indicated in brackets. The thick FM layer is the
spin-polarizer. Its fixed saturation magnetization
$\mathbf{M}_{s,p}$ points in the positive $z$-direction. The thin
FM layer is the spin-analyzer with saturation magnetization
$\mathbf{M}_{s,a}$. The direction of $\mathbf{M}_{s,a}$ is in
the plane of the layer. It can be parallel or anti-parallel to $\mathbf{M}%
_{s,p}$. Polarizer and analyzer are separated by a NM spacer of
thickness $t_{sp,1}\approx2$ nm. This is much smaller than the
spin-flip diffusion length (and any other scatter length) of a NM
spacer of a spin valve, so that spin is preserved between
polarizer and analyzer. The analyzer, in turn, is separated from
the insulating layer by a NM spacer of thickness $t_{sp,2}$
($t_{sp,2}\approx t_{sp,1}$). The different scattering experienced
by the minority- and majority-spin channels in the FM layers is
reflected in different resistivities $\rho_{FM}^{\min}$ and
$\rho_{FM}^{maj}$ , which obey $\rho_{NM}<<$
$\rho_{FM}^{maj}<\rho_{FM}^{\min}$ ($\rho_{NM}$ is the NM
resistivity). Outside the spin valve the left electrode is
continued with NM, which is also the material inside the orifice
and of the right electrode. The elastic mean free path of the
electrons, both in the NM and the FM, is supposed small compared
to the size of the orifice, so that transport is diffusive on this
scale.

The device is biased in the positive direction at a voltage $V$
($V>0$), applying $-V/2$ to the left electrode or reservoir and
$+V/2$ to the right reservoir. The resulting electron current
flowing from the polarizer to the orifice is spin-polarized. In
zeroth order, \textit{i.e. }without inelastic processes (inelastic
diffusion length exceeds orifice diameter), the resulting
spin-dependent electron-distribution function in the plane of the
orifice is
\begin{equation}
f_{\mathbf{k},\sigma}=\frac{1}{2}\left\{  \left[  1-\alpha\left(  \frac{1}%
{2}+\sigma\right)  \right]  f_{0}\left(  \epsilon_{\mathbf{k},\sigma}%
+\frac{eV}{2}\right)  +\left[  1+\alpha\left(  \frac{1}{2}+\sigma\right)
\right]  f_{0}\left(  \epsilon_{\mathbf{k},\sigma}-\frac{eV}{2}\right)
\right\}
\end{equation}
In Eq. (1) the degree of the spin-polarization induced by the polarizer is
given by $\alpha=\Delta R_{p}/R_{M}=(\rho_{FM}^{\min}-\rho_{FM}^{maj}%
)t_{p}/(\pi a^{2}R_{M})$. This is the difference in polarizer
resistance seen by minority-spin and majority-spin electrons,
normalized to the Maxwell resistance $R_{M}$, which is the
resistance of the corresponding diffusive, homogeneous NM point
contact. $\sigma$ denotes the electron spin in the polarizer
($\sigma=+1/2$ for majority-spin electrons and $\sigma=-1/2$ for
minority-spin electrons, defined with respect to
$\mathbf{M}_{s,p}$). Further, $f_{0}$ is the Fermi-Dirac
distribution, $\epsilon_{\mathbf{k},\sigma}$ is the total energy
of an electron in state $\mathbf{k}$ and with spin $\sigma$,
\textit{i.e.} inclusive the electrostatic energy, and $e$ is the
elementary charge ($e>0$). The distribution in the analyzer, since
it is so close to the orifice, to a good approximation is also
given by Eq. (1). The non-equilibrium distribution is similar to
that of a homogeneous diffusive NM point contact\cite{Kulik and
Yanson} or mesoscopic diffusive NM wires\cite{Pothier}. It is the
average of two Fermi step functions displaced with respect to each
other by an energy eV, which is the difference of the chemical
potentials of the reservoirs. In this case, however, the weight of
the functions is spin-dependent, so that for the energy range of
values intermediate between $f_{\mathbf{k},\sigma}=1$ and
$f_{\mathbf{k},\sigma}=0$ two values exist:
$f_{\mathbf{k},+1/2}=(1+\alpha)/2$ and $f_{\mathbf{k},-1/2}=1/2$ ,
for majority- and minority-spin electrons, respectively . To
arrive at Eq. (1), we assume that only the potential drop across
the polarizer, because of its large thickness and high
resistivity, is responsible for the modification of the electron
distribution with respect to the one corresponding to the
homogeneous NM point contact (which would correspond to $\alpha =
0$ in Eq.1).  Thus the effect of the analyzer is assumed
negligible. Further, we concentrate on the spin-dependent
contribution of the polarizer to the electron distribution of
Eq.1, neglecting the average over the electron spins.

Analogous to point-contact spectroscopy of magnons\cite{Kulik and
Shekhter}, \cite{Akimenko}, the electron distribution prepared in
a biased spin-valve point contact enables magnon emission by
electrons in the analyzer, up to a maximum magnon energy $eV$.
Relaxation of created magnons is dominated by absorption by
electrons . As usual for ferromagnets, we assume that electronic
transport is primarily due to $sp$-electrons, so that these
electrons control the magnon distribution, irrespective the
strength of electron-magnon coupling. We further assume that the
analyzer region exposed to the current is single domain. Finally,
at the present level, we neglect escape of created magnons from
the region exposed to the current and corrections to the
distribution given by Eq. (1) due to strong magnon absorption by
electrons. Thus, applying Fermi's golden rule to magnon emission
and absorption, and carrying out integrations over initial and
final states in spherical coordinates, we find the rate equation
for the occupation number of magnons $N_{\omega}$ with energy
$\hslash\omega_{\mathbf{q}}$ in the analyzer:

\begin{eqnarray}\label{align}
\frac{dN_{\omega}}{dt}    =\frac{1}{2\pi\hslash}\int d\epsilon
D(\epsilon )\int d\epsilon^{\prime}D(\epsilon^{\prime})\left|
\widetilde{g}\right| ^{2}[f_{\epsilon,\sigma}\left(
1-f_{\epsilon^{\prime},-\sigma}\right)
(1+N_{\omega})\delta(\epsilon-\epsilon^{\prime}-\hslash\omega_{\mathbf{q}%
})\allowbreak\\
  -f_{\epsilon^{\prime},-\sigma}\left(
1-f_{\epsilon,\sigma}\right)
N_{\omega}\delta(\epsilon^{\prime}-\epsilon+\hslash\omega_{\mathbf{q}%
})].\nonumber
\end{eqnarray}

Here $D(\epsilon)$ is the electron density of states normalized
with respect to the unit cell,
$\widetilde{g}$\ is an effective matrix element for
electron-magnon coupling, \textit{i.e.} renormalized with respect
to wave-vector-nonconserving scattering processes (see Appendix),
and the distribution $f$ is given by Eq. (2). Note that the
orientation of $\sigma$ in Eq.\ref{align} corresponds to minority
spins in the analyzer. The first term of the integrand applies to
emission, the factor $(1+N_{\omega})$ denoting the sum of
spontaneous and stimulated processes. By nature, this magnon
emission is incoherent, so that incoherent magnetization
precessions result. This is in contrast to the current-induced
coherent magnetization precession described in Ref. \cite{Slonc1}.
Magnons are spin unity quanta. According to spin conservation,
magnon emission thus is a spin-flip process, which must increase
the net spin of the electron system by unity by converting
analyzer minority-spin electrons into majority-spin electrons.
Accordingly, the first energy integration in Eq. (2) involves
minority-spin electrons, while the second integration involves
majority-spin electrons ($\epsilon$ and $\epsilon^{\prime}$\ no
longer depend on the wave-vector, as angular integrations were
done already). The second term of the integrand applies to
absorption, so that the role of minorities and majorities is
reversed. When $\mathbf{M}_{s,p}$ and $\mathbf{M}_{s,a}$ are
antiparallel, a minority-spin current in the analyzer was a
majority-spin current in the polarizer, so that for this
configuration $\sigma=+1/2$ in Eq. (1) gives the minority
distribution in the analyzer and $\sigma=-1/2$ the majority
distribution. For the parallel configuration, minorities and
majorities conserve their character when they enter the analyzer,
so that in Eq. (1) the usual sign convention applies.

To evaluate Eq. (2) for $T=0$, where the energy of created magnons
is limited to $\hslash\omega_{\mathbf{q}}<eV$ , one has to
recognize the possible emission and absorption processes in the
analyzer and to take into account the energy range and
distribution function (together termed ''phase volume'') of the
electron states involved.  Magnon emission takes an initial state
with $f_{\epsilon,+1/2}=(1+\alpha)/2$ to a final state with
$f_{\epsilon^{\prime
},-1/2}=1/2$, the phase volume being $(1+\alpha)(eV-\hslash\omega_{\mathbf{q}%
})/4$. For absorption there is a complementary process from a state with
$f_{\epsilon^{\prime},-1/2}=1/2$ to a state with $f_{\epsilon,+1/2}%
=(1+\alpha)/2$, with a phase volume $(1-\alpha)(eV-\hslash\omega_{\mathbf{q}%
})/4$. The sum of these contributions is $(\alpha/2)(eV -
\hslash\omega)$; the corresponding term give rise to a net
emission  magnons (which corresponds to our choice of the sign of
$\alpha$). In addition there are absorption processes from states
with $f_{\epsilon^{\prime},-1/2}=1$ to states with
$f_{\epsilon,+1/2}=(1+\alpha)/2$ and from states with
$f_{\epsilon^{\prime},-1/2}=1/2$ to states with
$f_{\epsilon,+1/2}=0$, with phase volumes
$\hslash\omega_{\mathbf{q}}/2$ and
$(1-\alpha)\hslash\omega_{\mathbf{q}}/2$, respectively. The total
phase volume is $\hslash \omega(1 - \alpha/2)$. Note that this
contribution to absorption exists for purely equilibrium state of
the electron system ($V = 0, \alpha = 0$). In the parallel
configuration, \textit{mutatis mutandis}, similar processes with
similar phase volumes occur. For either configuration Eq. (2) thus
goes over into
\begin{equation}\label{3}
\frac{dN_{\omega}}{dt}=-\frac{1}{\tau_{m-e}}\left[  N_{\omega}\left(
1+\frac{eV}{\hslash\omega_{\mathbf{q}}}S_{z}\right)  -\frac{eV-\hslash
\omega_{\mathbf{q}}}{4\hslash\omega_{\mathbf{q}}}\left(  1-2S_{z}\right)
\right]  .
\end{equation}
$S_{z}=\alpha(\mathbf{M}_{s,a}\cdot\mathbf{M}_{s,p}/2M_{s,a}M_{s,p})$ is a
projection of $\mathbf{M}_{s,a}$ on $\mathbf{M}_{s,p}$, weighed by the current
polarization. Whereas the spin polarization is defined with respect to
$\mathbf{M}_{s,p}$, $S_{z}$ takes into account the sensitivity of
magnon-electron processes to the polarization with respect to the direction of
$\mathbf{M}_{s,a}$. In Eq. (3) $\tau_{m-e}\approx\hslash^{-1}\left|
\widetilde{g}\right|  ^{2}\left[  D(\epsilon_{F})\right]  ^{2}\hslash
\omega_{\mathbf{q}}$\ is the characteristic time for magnon-electron
processes, $D(\epsilon_{F})$ being the electron density of states at the Fermi level.

Eqs. (2) and (3) describe transfer of spin from the spin-polarized current to
the magnon system. Due to spin transfer the population of non-equilibrium
magnons increases until a steady state is reached where magnon emission and
absorption balance each other, \textit{i.e} where $dN_{\omega}/dt=0$. For
magnons of energy $\hslash\omega_{\mathbf{q}}$ the steady state is
characterized by an effective magnon temperature $T_{m,\omega}^{eff}$,
obtained by equating the number of such magnons to the average population as
given by the Planck distribution ($k_{B}$ is Boltzmann's constant):
\begin{equation}\label{4}
T_{m,\omega}^{eff}=\frac{1}{k_{B}}\frac{eV-\hslash\omega_{\mathbf{q}}}{4}%
\frac{1-2S_{z}}{1+(eV/\hslash\omega_{\mathbf{q}})S_{z}}.
\end{equation}
In the limit of weak polarization and for magnon energies $\hslash
\omega_{\mathbf{q}}<<eV$ one obtains $T_{m,\omega}^{eff}\approx
eV/4k_{B}$. The effective temperature is larger for $S_{z}<0$
(antiparallel configuration) than for $S_{z}<0$ (parallel
configuration). This is natural, because the phase volume for
magnon creation processes is larger when $S_{z}<0$. Moreover, in
the antiparallel configuration $T_{m,\omega}^{eff}$ diverges at a
critical voltage given by
\begin{equation}\label{cri}
eV_{c}S_{z}=-\hslash\omega_{\mathbf{q}}
\end{equation}
 This is interpreted as an unlimited increase of the number of
magnons due to stimulated emission, resembling a magnon avalanche.
This highly excited state of the analyzer goes along with a
strongly suppressed magnetization and may lead to a kind of
critical behavior with similarities to the phase transition to the
normal state at the Curie temperature, which results from strong
thermal excitation of magnons.  As
seen in Eq. (4), at the voltages larger than the critical one
$dN_{\omega}/dt$ is positive so that $N_{\omega}$ has a positive
increment.

Let us introduce the magnon concentration $n_m$ normalized with respect to elementary cell
$$ n_m = \int {\rm d} \omega \nu_m N_{\omega} $$
where $\nu_m(\omega) $ is magnon density of states normalized by elementary cell volume. It is clear that $n_m = 1$ would correspond to complete suppression of magnetization. Note that a decrease of total magnetization of the ferromagnet
$|\bf M|$ with an increase of $n_m$ for small $n_m$ is a
well-established fact and describes in particular a decrease of
$|\bf M|$ with temperature increase. This behavior can be
interpreted as a result of uncertainty of the orientation of $\bf
M_{s,a}$ within the angle $\theta \sim n_m$ with respect to its
orientation at $n_m = 0$.  As for the situation of very high magnon occupation numbers when $n_m \sim 1$, to the best of our knowledge
it still has not met a consequent theoretical treatment.
One could speculate that the magnon avalanche leading to
$n_m \sim 1$ corresponds to
switching since for $n_m \simeq 1$ the average, time-independent
magnetization of the sample would be completely suppressed. At the
same time any local fluctuation with the with the opposite
direction of magnetization would be magnified due to the fact that
the sign of the projection of the incoming spins on this direction
entering the driving terms in Eq.\ref{3} would correspond to
magnon absorption rather than to magnon emission.

However one can expect that
the values  of
$n_m$ can be stabilized at some $n_m << 1$ due to mechanisms not
included in
Eq. (4). To look for such mechanisms we should consider magnon
kinetics at large occupation numbers (but still within the framework of
conventional magnon theory implying $n_m << 1$.

\section{Role of magnon-electron processes}

First let us discuss a possible role of magnon-electron interactions.
So far we assumed that the electron distribution given by Eq.1
exists even at high magnon concentration. However one can expect
that a high magnon emission rate leads to decay in the analyzer of
the spin-dependent part of the electron distribution. The latter
consequently can no longer support further strong emission. To
understand the evolution of this distribution with an increase of
the magnon concentration we make use of the electron-magnon
collision operator:
\begin{eqnarray}\label{em}
I_{e-m} = \int{\rm d} \omega \nu_{\omega} |{\tilde g}|^2 (
f_{\varepsilon,\sigma}(1 - f_{\varepsilon',-\sigma})\lbrack (1 +
N_{\omega})\delta(\varepsilon - \varepsilon' - \hslash \omega) +
N_{\omega}\delta(\varepsilon' - \varepsilon + \hslash
\omega)\rbrack \nonumber \\
-f_{\varepsilon',-\sigma}(1 - f_{\varepsilon,\sigma})\lbrack (1 +
N_{\omega})\delta(\varepsilon' - \varepsilon - \hslash \omega) +
N_{\omega}\delta(\varepsilon - \varepsilon' + \hslash
\omega)\rbrack )
\end{eqnarray}
Here $\nu(\omega)$ is the magnon density of states. As it is seen,
for $N_{\omega} >> 1$ with a neglect of the spontaneous processes
this operator tends to establish an electron distribution of the
sort $f_{min}(\varepsilon) = f_{maj}(\varepsilon - \hslash
\omega)$. In addition, the electron-magnon processes even at very
large $N_{\omega}$ can only take place between of the electronic
majority and minority states separated by the energy $\hslash
\omega$ since the minority electron arising as a result of magnon
absorbtion by the majority electron can not further absorb magnons
and vice versa, the minority electron turning to the majority
electron by an emission of magnon can not further emit magnons.
Having these facts in mind one sees that, in particular, the
integral of Eq.\ref{em} is cancelled by the electron distribution
\begin{eqnarray}\label{satp}
f_{maj} = \frac{1}{2}\left( f_0 (\varepsilon + \frac{eV}{2} +
\hslash \omega) + f_0(\varepsilon - \frac{eV}{2})\right) \nonumber
\\
f_{maj} = \frac{1}{2}\left( f_0 (\varepsilon + \frac{eV}{2})  +
f_0(\varepsilon - \frac{eV}{2}- \hslash \omega)\right)
\end{eqnarray}
So the electron-magnon coupling does not lead to a total decay of
spin polarization, however the total density of polarized spins
supported by the distribution resulting from the coupling in
question is $(\alpha/2)\hslash\omega D$ while the initial electron
distribution given by Eq.[1] corresponds to the density of spins
$\alpha eV/2 D$. So one concludes that for $V
> V_c$ the electron-magnon coupling within the polarizer whatever
strong it is can not completely suppress the spin polarization.

\section{Role of magnon-magnon processes}

Now let us discuss a possible role of magnon-magnon processes
which can become important at high phonon occupation numbers. Note
that these processes can be considered as a precursor of the
general nonlinearity of the magnon physics at high magnon
excitation levels.
 One discriminates between 3-magnon processes (originating
due dipolar-dipolar interactions and non-conserving total spin)
and 4-magnon processes related to purely exchange interactions and
conserving total spin and total number of magnons (see e.g.
\cite{Akhiezer}). For the simplicity we will consider purely 3D
magnon spectrum ($\nu_m(q) \propto q^2$) while generalization for
the case of more complex spectrum is straigtforward). Let us start
from 3-magnon processes. For the convolution of 2 magnons of the
modes $1$ and $2$ into a magnon of the mode $3$ the efficiency
rate can be estimated as \cite{Akhiezer}
\begin{equation}\label{3m}
\frac{1}{\tau_3} \sim
\frac{2\pi}{\hbar}\frac{(\mu_B M)^2}{\Theta_C}
\gamma_3(1,2)N_{2}(N_{3}+ 1)
\end{equation}
where $\Theta_C$ is the Curie temperature,  $M$ is magnetic moment
density while $\gamma_3(1,2)$ is a dimensionless parameter
depending on the details of magnon spectrum.  Note that the
parameter $\gamma$ crucially depends on the details of magnon
spectrum and on the momentum conservation law. For standard 3D
magnon spectrum $\gamma \sim q_2a$ where $q_2$ is the wave vector
of the mode $2$. However if $\omega(q \rightarrow 0)$ does not
depend on the direction of $\bf q$ momentum conservation law
allows coalescence processes only if $\omega_3 > 3 \omega(q
\rightarrow 0$ (see e.g. \cite{Akhiezer}. However in general case
\begin{equation}\label{spectrum}
\omega_q^2 = (\mu_B H_i + 2 JS(qa)^2)(\mu_B H_i + 2 JS(qa)^2 + 4
\pi \mu_B M \sin^2\theta_{\bf q})
\end{equation}
where $H_i = H_0 - 4\pi\mu_BM + H_a$, $\theta_{\bf q} = \angle
({\bf q},{\bf M})$ while $H_0$ and $H_a$ are external magnetic
field and anisotropy field, respectively. As it is seen, in this
case the restrictions related to momentum coservation are not as
severe due to a presence of the term depending on $\theta_{\bf
q}$. One also notes that while the magnon-magnon processes in the
bulk of a perfect crystal definitely obey momentum conservation
law,  it is not necessarily the case for imperfect systems – as it
is shown in particular in the Appendix for electron-magnon
interactions. One can expect that the efficiency of processes
violating the momentum conservation can be as high as the for
non-violating ones if the spatial scale of inhomogeneity is
comparable with magnon wavelength.

In its turn, the rate of 4-magnon process $(1,2 \rightarrow 3,4)$
can be estimated as
\begin{equation}
\frac{1}{\tau_4} \sim \gamma_4(1,2,3) N_2(1+N_3)(1+N_4)
\end{equation}
where $\gamma_4(1,2,3)$ is the corresponding dimensionless parameter which
for standard 3D spectrum and wave vectors of the same order of magnitude
is equal to $(qa)^8$.
One
notes that since 4-magnon processes simultaneously conserve the
total number of magnons and the total energy, they can not
efficiently modify the magnon distribution if it is initially
concentrated at lowest possible energies (ensuring the largest
gain for the magnon avalanche). So we will mainly concentrate on
3-magnon processes. The magnon occupation numbers $N$ can be
written in terms of our normalized parameters $n_m$ as
\begin{equation}
N_{\omega} \sim n_{m,\omega} \omega {\cal V}_{\omega}^{-1}
\end{equation}
where $\cal V$ is the phase volume available for the corresponding
modes (for standard 3D spectrum $\omega ({\bf q})$ ${\cal V} \sim
(qa)^3$). So one concludes that despite of relatively small
efficiency of dipolar-dipolar interactions (related to small
matrix element $\mu_B M$), it can be of importance for small
magnon wave vectors where the corresponding smallness can be
compensated – even at $n_m < 1$ - by large factors
$V_{\omega}^{-1}$ resulting from the phase volume considerations.

First let us consider a situation when the occupation numbers of
the resulting magnons $N_3$ are small. In this case one can
rewrite Eq.\ref{3} with an account of 3-magnon processes as
\begin{equation}\label{11}
\frac{dN_{\omega}}{dt}=-\frac{1}{\tau_{m-e} \hbar \omega}(S_zeV +
\hbar \omega)N_{\omega} -
 \frac{2\pi}{\hbar}\frac{(\mu_B M)^2}{\Theta_C} n_m
 \frac{\gamma_3}{{\cal V}_{\omega}}
 N_{\omega}
\end{equation}
So one sees that
the magnon avalanche is stabilized at
\begin{equation}
{\tilde n_m} = (|S_zeV| - \hbar \omega)\frac{\hbar
\Theta_C}{2\pi(\mu_B M)^2)\tau_{m,e}\hbar \omega}\frac{{\cal
V}_{\omega} }{\gamma_3}
\end{equation}
One notes that in this
case the gradual evolution of $n_m$ is possible
with an increase of $V$ (starting from $V_c$) following by the
reversal of magnetization at some $V = V_{c1}$
corresponding to $n_m(V) \sim 1$. As for the numerical estimates, one
notes that for typical ferromagnets $\Theta_C/2\pi \mu_B M \sim
10^3$. Then, according to the estimates given in Appendix
$\tau_{m,e}\omega \sim 10 - 100$. Thus, assuming $\gamma_3 \sim 10^{-2}$
one obtains
$$ n_m \sim (10^{-3} -  10^{-2})\frac{(|S_zeV| - \hbar
\omega)}{\hbar \omega}\frac{\hbar\omega}{\mu_BM} $$
where $\mu_BM
\sim 0.01 mV$. Thus the values $n_m < 1$ are still expected for
$|V - V_c| \sim 1 mV$
However it may occur that for small values of $\gamma_3$
$V_{c1}$ is too close to $V_c$ to allow a wide region for the gradual evolution.

Now let us consider a situation
when the occupation numbers $N_3$ can be large which lead to stimulated
emission of the corresponding magnons.

For the mode $\omega_3$ one has
\begin{eqnarray}
\frac{dN_{\omega_3}}{dt}=-\frac{1}{\tau_{m-e}\hbar \omega_3}\left(
(S_zeV + \hbar \omega_3)N_{\omega_3} - \frac{eV-\hslash
\omega_3}{4\hslash\omega_3}(  1-2S_{z})\right) + \nonumber\\
\frac{2\pi}{\hbar}\frac{(\mu_B M)^2}{\Theta_C}
\frac{\gamma_3}{{\cal V}_{\omega}^2} n_m^2 (N_{\omega_3} +1)
\end{eqnarray}
We will assume that, since $\omega_3$ is at least twice larger
than $\omega_1$,  $ S_zeV + \hbar \omega_3 > 0$.

So in stationary situation one has
\begin{equation}
N_{\omega_3} \sim \frac{n_m^2 + \beta }{n_c^2  -  n_m^2}
\end{equation}
where
\begin{equation}
n_c^2 = \frac{1}{\tau_{m-e}\hbar \omega_3}(S_zeV + \hbar \omega_3)
\left( \frac{2\pi}{\hbar}\frac{(\mu_B
M)^2}{\Theta_C}\frac{\gamma_3}{{\cal V}_{\omega}^2}
 \right)^{-1}
\end{equation}
while
\begin{equation}
\beta = \frac{eV-\hslash \omega_3}{4 \tau_{m,e}\hslash\omega_3}(
1-2S_{z}) \left(\frac{2\pi}{\hbar}\frac{(\mu_B M)^2}{\Theta_C}
\frac{\gamma_3}{{\cal V}_{\omega}^2}(N_{\omega_3} +1) \right)^{-1}
\end{equation}
So if
$\beta < n_m^2$ the stationary solution for $N_{\omega}$
corresponds to
\begin{equation}
(|S_zeV| - \hbar \omega)\frac{\hbar \Theta_C \frac{{\cal
V}}{\gamma_3}}{2\pi(\mu_B M)^2)\tau_{m,e}\omega}(1 -
\frac{n_m^2}{n_c^2}) = n_m
\end{equation}
or
\begin{equation}
n_m = - \frac{n_c^2}{2{\tilde n_m}} +  \frac{n_c^2}{2\tilde n_m}
 \left( 1  +
\frac{4\tilde n_m^2}{n_c^2} \right)^{1/2}
\end{equation}
As it is seen, the behavior depends on the ratio
\begin{equation}
\frac{n_c}{\tilde n_m} = N_{\omega_3}^{-1} \sim \frac{(S_zeV +
\hbar \omega_3)^{1/2}}{(|S_zeV - \hbar \omega)|}\frac{\mu_B
M}{\Theta_c^{1/2}(\tau_{m-e}\omega)^{1/2}} \gamma_3^{-1/2}
\end{equation}
Namely, in the first scenario ${\tilde n_m} < n_c$
\begin{equation}
n_m \sim {\tilde n_m}
\end{equation}
In this case the gradual evolution of $n_m$ is possible provided
$V_{c1} - V_c$ is large enough. In the second scenario ${\tilde
n_m}  > n_c$ one has
\begin{equation}
n_m \sim n_c
\end{equation}
 While at $V - V_c \rightarrow 0$ one always has
$n_m \sim {\tilde n_m}$, at higher $V$ the condition ${\tilde n_m}
> n_c$ can be reached at some $V = V_{c2}$ provided $V_{c2} <
V_{c1}$. It is important to note that when $n_m \sim n_c$  the
value of $n_m$ {\it decreases} with an increase of $|V|$ since the
magnon generation corresponds to $S_zeV < 0$.

Thus, the non-monotonous evolution of magnetization is expected:
at critical bias $S_zeV = - \hbar \omega $ the value of $n_m =
n_c(V=V_c) $ is reached while the further increase of $V$ leads to
a decrease of the magnons number. One notes that in this regime
$$ N_{\omega_3} \sim \frac{\tilde n_m}{n_c} $$
and thus $N_{\omega_3}$ increases with the increase of $V$. The
situation again becomes "critical" when the values of $N_3$
becomes comparable with $N$ and a situation of energy diffusion is
established. However we will not discuss this situation in more
detail.

Note that the actual physical picture with an account of magnon-magnon
processes is rather complex and sensitive to details of magnon spectrum etc.
Thus our simplified analysis aims only to reveal the main features.
We can conclude that the magnon-magnon processes can lead to
saturation of magnon emissions at $n_m < 1$, that when no switching occurs.
In this case the further gradual evolution of magnetization
with an increase of $V$ is possible. The scenario depends on the relation between
$n_c$ and $\tilde n_m$.

\section{Comparison to previous theoretical approaches}

A short comparison with the approach of Slonczewski\cite{Slonc1}
is in place. In the latter case the spin transfer from the
incident electrons to the layer is proportional to the amplitude
of magnetization precession and is zero for $\bf M$ being parallel
or antiparallel to $\bf M_0$ while in our case the spontaneous
processes exists as well allowing spin relaxation even for the
spins parallel or antiparallel to $\bf M_0$. In the case of
mechanism by Slonczewskii the spin transfer from the incident
electrons is equal to $j_{s,inj} \sin \theta $ provided $d > l_{p}
\sim (\hbar/p_F)\varepsilon_F/E_{ex}$ where $E_{ex}  $ is the
exchange energy, $l_p$ is the spin precession length and
$j_{s,inj}$ is the spin current. Since $E_{ex}$ is not too small
with respect to the Fermi energy this condition holds for
practical values of $t_a$. Thus this mechanism is restricted by
the near-surface layer with a thickness $l_p$. The evolution of
the precession angle is described by Eq.17, the key equation of
the paper, which can be in our terms be written as
\begin{equation}
\frac{{\rm d}\theta}{{\rm d}t} = - (\alpha \omega - \frac{j_{s,inj}}{t_a})
\sin \theta
\end{equation}
where $\alpha$ is the Gilbert parameter while $j_{s,inj}$ is
the spin current in the analyzer. (Note, that, strictly
speaking, the applicability  of Gilbert damping, suggesting the
equilibrium state of the ferromagnet, is questionable for the
non-equilibtium electron distribution). To compare with our
considerations we first note that for small $\theta$ the number of
"coherently excited magnons" can be related to $\theta$ as $n_m
\sim \theta^2/2$. Then, one can consider the product
$\alpha\omega$ as the magnon relaxation rate, $1/\tau_m$. We
believe that the dominant mechanism of the magnon relaxation is
related to magnon-electron coupling discussed above.  Thus one can
rewrite this equation as
\begin{equation}\label{slonlike}
\frac{{\rm d}n_m}{{\rm d}t} = (\frac{j_{s,inj}}{t_a} -
\frac{1}{\tau_m})n_m
\end{equation}
As it is seen,  the criterion for the current-driven evolution of magnetization
can be written for the Slonczewskii mechanism as
\begin{equation}\label{critslon}
 j_{s,inj} > \frac{t_a}{\tau_m)}
\end{equation}
which can be rewritten in terms of the bias with an account of the
estimate of $\tau_m$ given above as
\begin{equation}
V > \eta^{-1} V_c
\end{equation}
where $ \eta  = j_{s,inj}/j_{s,inj}^{ball}$ while
$j_{s,inj}^{ball}$ is the spin current which would exist
under the same bias provided the structure would be ballistic.
One notes that this criterion differs from that given by
Eq.\ref{cri} by a factor $\eta^{-1} > 1$ in the r.h.s. of the
inequality which for the diffusive electron transport is much
larger than unity. Thus there is a broad region of biases
$$ \hbar_{\omega} < |S_z|eV < \eta^{-1} \hbar \omega  $$
where the excitation of magnons according to our scenario is
possible  while the mechanism by \cite{Slonc1}  is not efficient.

Note however that the approach suggested in our paper is
completely self-consistent one and is limited by
quantum-mechanical treatment while the approach of \cite{Slonc1}
exploits a combination of quantum-mechanical treatment of electron
spin evolution and semi-classical transfer of $x-$ component of
electron spin to the layer as a whole described with a help of
classical Landau-Lifshits equation. As a result, some questions do
not meet unequivocal treatment - in particular a mechanism leading
to a change of $z-$component of spin by portions equal (1/2) which
governs the spin evolution in a consequent quantum-mechanical
picture and is expected to have "non-coherent" character. It is
also not clear in what way the single-electron spin is transferred
to the excited region as a whole since any given electron is
coupled only to its closest surrounding.

To some extent these questions were noted by Berger \cite{Berger}
who exploited the similar mechanism of boundary-induced spin
transfer to describe emission of spin waves by incoming electron
spin flux. He attempted to write the equation for $z-$component of
electron spin, but he simply postulated  it in the most simple
relaxation time approximation form to fit a quantum-mechanical
principle that $z$-component of spin can be changed by portions of
$1/2$ (Eqs. (9-12) in \cite{Berger}). In this way he also came to
a conclusion concerning instability of magnon system for the
current exceeding some threshold value similar to one we have
formulated on the base of our consistent quantum-mechanical
treatment. Note however that, as the paper \cite{Slonc1}, this
approach does not take into account a possibility of
non-equilibrium electron distribution, relating the chemical
potential difference majority and minority subbands to a
difference in pumping of electron momenta at the boundary due to a
difference in drift currents. To our opinion such an approach is
equivalent to an assumption that the energy diffusion length is
equal to elastic mean free path and strongly underestimates the
real potential difference.

Since one notes similarities between our results and results of
\cite{Slonc1}, \cite{Berger} including a possibility of switching
discussed above, magnon escape factor etc. one can consider our
scheme as a generalization of approaches by Slonzcevskii and
Berger. But we also predict a possibility of stabilization in
magnon system at the levels still much less than corresponding to
switching - in contrast to results by \cite{Slonc1},
\cite{Berger}.

As for high excitation levels $V - V_c \geq 1$ we would like to
note that there is still a problem whether the real magnetization
evolution is in accordance to our "incoherent" scenario or it has
features of "coherent" classical magnetization evolution of the
sort considered in \cite{Slonc1} (with some correction of the
Gilbert parameter $\alpha$ for the nonequilibrium case) . In any
case one expects that if initially $\bf M $ and $\bf M_0$ are
parallel or antiparallel an initial stage is described by our
scenario since it is  related to magnon excitation which can be
considered as quantum fluctuations and exist even for $\theta =
0$. The further behavior is not as clear since, as we have noted
above, the situation of very high magnon occupation numbers is
still not perfectly clear. In particular one can expect a sort of
"producing of coherency" originated due to effective magnon-magnon
processes and establishing the "coherent" evolution. Note that a
similar problem was discussed for acoustoelectric generators of
sound waves where the coherent signal arose (or not arose) from
initial electron-drift driven emission of incoherent phonons
(\cite{Katil}). Certainly this choice can crucially dependent on
the factor of boundaries which can support - or destroy - such
coherency, and on the excitation level (that is on the product
$|S_z|eV$).

We would like also to emphasize that for non-ballistic limit where
the electron-magnon coupling in our case has a bulk character (see
Appendix) in contrast to mechanisms of spin transfer considered in
\cite{Slonc1}, \cite{Berger} our "bulk" mechanism is expected to
dominate in any case.

To conclude this Section, we would like in addition to consider a
situation when the exciting current passes only through small area
of the analyzing layer which takes place when the planar
multilayered structure is excited with a help of a point contact
attached to such a structure. This geometry corresponds in
particular to experimental situation of the papers \cite{Tsoi1},
\cite{Tsoi2} which will be to some extent discussed in what
follows. In contrast to the picture suggested above, the emitted
magnons have a possibility to escape the excited region
propagating along the plane of the analyzing layer (assumed to
have an infinite area). If one demotes a size of the excited
region (being equal to the size of the point contact) as $a$, the
escape time $\tau^{esc}_m$ is expected to be equal to $a/v_m$
where $v_m = hk_m/m_m$; here $v_m$ is a magnon velocity while
$k_m$ is the magnon wave vector with a smallest possible value of
$\sim 1/a$. Correspondingly, the largest possible value for
$\tau^{esc}_m$ is obviously $a^2m_m/\hbar$.

This escape can be taken into account by adding to Eq. (3) the
relaxation term $-N_{\omega}/\tau_{m}^{esc}$, where
$\tau_{m}^{esc}$ is the
magnon escape time for the point-contact geometry. This gives $\tau_{m-e}%
/\tau_{m}^{esc}$ as extra term in the denominator of Eq. (4).
Divergence of $T_{m,\omega}^{eff}$ is not prevented by magnon
escape, but fast escape $(\tau_{m-e}/\tau_{m}^{esc}>>1)$ leads to
a strong increase of the critical
voltage according to
\begin{equation}\label{vc}
eV_{c}S_{z}\approx-(\tau_{m-e}/\tau_{m}^{esc}%
)\hslash\omega_{\mathbf{q}}
\end{equation}
 Taking into account that
$\tau_{m-e}^{-1} \simeq \omega /(k_Fd)$ one obtains
\begin{equation}\label{rtau}
 \tau_{m-e}/
\tau_m^{esc} \geq \frac{k_F d}{\hbar \omega}\frac{\hbar^2}{a^2
m_m}
\end{equation}
 where the equality corresponds to the smallest magnon
in-plane wave vector $\sim 1/a$. As it is seen, in this case the
criticality starts at the lowest possible in-plane wave vector. So
the criticality criterion is in principle independent of the
magnon frequency provided the ratio given by Eq.\ref{rtau} is
larger than unity.

Note that according to Eq.\ref{vc} in the corresponding regime
$V_{c}\varpropto1/(a^2)$. For diffusive electron transport the
critical current is $I_{c}=V_{c}/R_{M}$ is thus independent of the
orifice size which agrees with prediction by Slonczewski for the
similar geometry although in terms of coherent excitation of spin
waves \cite{Slonczewski2}

\section{Comparison to experimental situation}.

Thus we have shown that the incoherent stimulated emission of
magnons trigger an avalanche evolution of magnetization at biases
smaller than ones necessary for the switching behavior according
to the mechanism suggested by Slonczewski. However the final stage
at these moderate biases corresponds to the state with very high
magnon temperature but still with the initial direction of
magnetization. Such a behavior is different from "switching"
predicted in \cite{Slonc1}. In our scenario the first step-like
evolution of magnetization with the applied bias (at $V = V_c =
|\hbar \omega_m/eS_z|$ can be followed by the gradual increase of
the number of magnons and, correspondingly, by a gradual decrease
of the $z$-component of total magnetization. Note that neither the
approach by Slonczewski nor the approach by Berger have predicted
such a gradual evolution. Unfortunately the actual physical
picture at this stage which is controlled by magnon-magnon
processes is rather complex and sensitive to details of magnon
spectrum etc. Thus our simplified analysis allows only to reveal
the main features and does not allow direct quantitative
comparison to experimental data.

First we note that qualitatively many experiments on
current-driven excitations in magnetic multilayers
\cite{Tsoi1},\cite{Myers}, \cite{Katine},\cite{Tsoi2},
\cite{Theeuwen} unambiguously exhibited signatures of a gradual
evolution of magnetic state with an increase of bias.

In the papers \cite{Myers}, \cite{Katine} current-induced
switching of magnetic moments in Co/Cu/Co pillar  structures was
reported. The switching exhibited hysteretic behavior as function
of injected current and external magnetic field and was
interpreted in terms of theory by Slonczewski \cite{Slonc1}. Note
however that the initial phase of the process clearly indicated a
gradual evolution with an increase of the current following
step-like increase of differential resistance. At high external
magnetic fields instead of switching the data exhibited spikes for
one of the directions of the current which were attributed to
emission of spin waves. Since the theory \cite{Slonc1} exploiting
a constant value of Gilbert parameter does not allow an existence
of stable spin-precessing mode, the authors of \cite{Katine}
explained the observed behavior as a result of a dependence of
$\alpha$ on the angle of precession. Although the latter mechanism
can in principle explain spike-like features, it can hardly
explains the gradual evolution. Note that the gradual evolution of
contact resistance with an increase of the current depending on
the current direction was also observed in \cite{Theeuwen}.

The transition from hysteretic behavior to the spike-like one with
an increase of magnetic field was also reported in \cite{Urazdin}
for permalloy-based nanopillars. We would like to note that
authors of \cite{Urazdin} have  suggested a possible role of
nonequilibrium magnons with high temperatures to explain the
behavior observed. The also observed a presence of telegraph noise
in the vicinity of the spike. It is important to note that the
noise was pronounced only at very narrow region of biases and has
not revealed any trace of hysteretic behavior. This fact could not
be easily understood suggesting magnetization reversal since in
the latter case the anisotropy field would give rise to a
hysteretic behavior.

To our opinion, the spike can be related to stabilization of the
magnon occupation numbers due to magnon-magnon scattering
according to the scenario (2) considered above. Indeed, for the
resulting "incoherent precession state" the direction of
magnetization is only slightly "smeared" with respect to its
original direction and thus no hysteretic behavior is expected. In
contrast, the variation of the bias around the corresponding
threshold value leads to a reversible behavior of the magnon
occupation number. The step-like increase of resistance leading to
the spike in differential resistance can be related to additional
resistance related to effective electron-magnon scattering rather
than to magnetization reversal.

Correspondingly, the transition from hysteretic behavior to
spike-like behavior can be related to the transition from scenario
(1) to scenario (2). Indeed, with an increase of magnetic field a
role of the angle-dependent term in \ref{spectrum} decreases with
respect to the term $\mu_BH_i$. As a result, the the phase volume
for 3-magnon processes decreases thus leading to a decrease of
$\gamma_3$. As it is seen, it is favorable for regime (2) where an
increase of bias does not lead to an increase of $n_m$ at least in
some interval of biases.

In the papers \cite{Tsoi1}, \cite{Tsoi2} magnetic mulitilayers wee
excited with a help of point contact. Correspondingly, the
excitation took place only at small area of the multilayer
concentrated around the point contact which is in agreement with
model discussed at the end of our Sec.2. The differential
resistance spectra obtained in \cite{Tsoi1} for Co/Cu multilayers
excited by the metallic point contact clearly demonstrated a
gradual increase of resistance following an initial spike which
the authors ascribed to stimulated emission of magnons according
to scenario \cite{Berger}.

To our opinion, the experimental data mentioned above evidence the incoherent character
of magnons emission (at least at the initial stage) since it is the latter mechanism
that can -
according to our model - explain co-existence of "critical" spike-like and"switching"
behavior with a gradual evolution.

\section{Conclusions}

To conclude, we have given a consequent quantum-mechanical
treatment of spin-polarized transport in spin-valve point contact
with an account of non-equilibrium electron distribution is given.
It is shown that at large biases an avalanche-like creation of
low-frequency magnons within the ferromagnetic layer up to is
possible which agrees with earlier predictions based on
semi-phenomenological models. However in contrast to earlier
studies \cite{Slonc1} it is found that the stabilization within
the magnon system can be achieved at finite (although large)
magnon temperatures and the gradual evolution of such a state with
the bias is possible. These results are in agreement with existing
experimental data.

\section{Appendix}

Let us consider an effect of boundaries and disorder on the
magnon-electron matrix elements.
Following Mills et al. \cite{Fert} we will
write the element corresponding to a transition
from the electron state $\bf k$ to the electron
state $\bf k'$ in a presence of created magnon in the state
$\bf q$ as
\begin{equation}\label{initial}
{\cal M}|_{\bf k \rightarrow k' } = J\frac{\bf s M}{|s||M|}
\frac{ a^{3/2}}{ V^{3/2}}\int{\rm d}{\bf r}^3 \exp i({\bf k - k' -
q}){\bf r}
\end{equation}
where $\bf s$ and $\bf M$ are the electron spin and sample
magnetization, $J$ is an exchange constant, $a$ is the lattice
constant, $\bf k, k'$ and $q$ are the wave vectors of initial and
final states of electrons and the magnon wave vector,
respectively, while $V$ is the normalizing volume. One notes that
typically the integration gives a standard momentum conservation
law. However now we are interested in a case when the momentum
conservation is violated which in particular is related to a
finite size of the layer with a thickness $t_a$ in $x$-direction.
Thus we will assume that in plane of the layer the momentum
conservation holds in a standard way and will concentrate on the
integration over $x$. The corresponding factor arising in the
expression for ${|\cal M|}^2$ can be readily written as
\begin{equation}\label{kx}
\frac{a}{{t_a}^3(k_x -k_x' - q_x)^2}
\end{equation}
For a given initial and final electron energies $\varepsilon
=\varepsilon_{\bf k, - \sigma }$ and $\varepsilon' =
\varepsilon_{\bf k',  \sigma }$ One notes that the Fermi surfaces
are typically separated by relatively large gap $\Delta k_F \simeq
k_F E_{ex}/\varepsilon_F$. For long-wave magnons we expect $q <<
\Delta k_F$ which allows us to neglect $q$ in our estimates. We
will first integrate over angular variables $\vartheta,
\vartheta'; \varphi, \varphi'$. Having in mind that the difference
$\varepsilon - \varepsilon'$ (controlled by the distribution
functions) is much less than $E_{ex}$ we also will neglect this
difference in course of the  angular integration. Thus one has
$k_x = k_{F,-}\cos\vartheta$ and $k_x' = k_{F,+} \cos \vartheta'$.
 The momentum conservation in perpendicular direction
eliminates integrations over $\varphi'$ and $\vartheta'$ with an
obvious result $\sin \vartheta' = (k_{F,-}/k_{F,+}) \sin \theta$
(which is related to a conservation of the in-plane component of
the momentum). Correspondingly, one obtains
$$\cos \vartheta' = \left(1 - (k_{F,-}/k_{F,+})^2 \sin^2 \vartheta
\right)^{1/2} $$
One clearly see a gap preventing small values of $\cos \vartheta'$
(since the modulus of $x$-component of electron wave
vector for majority electron is larger than for minority)
and finally obtains for the denominator in Eq.\ref{kx}:
\begin{equation}
 t_a^3 k_F^2 \left( \xi - \sqrt{ \xi^2(1 -
2\frac{\Delta k_F}{k_F})+ 2 \frac{\Delta k_F}{k_F}} \right)^2
\simeq t_a^3k_F^2 \left( \frac{\Delta k_F}{k_F}(\xi - \frac{1}{\xi^2}
\right)^2
\end{equation}
where $\xi = \cos \vartheta$.
Thus performing the integration over $\xi$ one has a factor
$$ \sim \frac{a}{t_a^3 \Delta k_F^2} $$
Integrating over angular variables
corresponding to the initial and final
states one notes that the effective matrix
element can be estimated as
\begin{equation}
\tilde{|\cal M|}^2 = J^2 (k_F/(\Delta k_F)^2t_a)
\end{equation}
The effective rate equation can be rewritten as
\begin{eqnarray}\label{rate1}
\frac{{\rm d} {\cal N}_{\omega }}{{\rm d} t} =
\frac{1}{2 \pi \hbar} \sum_{\sigma}
\int{\rm d}{\varepsilon} \nu (\varepsilon)
\int{\rm d} \varepsilon'\nu(\varepsilon')|{\tilde {\cal M}}|^2
( f_{\varepsilon,
\sigma}(1 - f_{\varepsilon',-\sigma})(1 + {\cal N_{\omega}})
\delta(\varepsilon - \varepsilon'+
\hbar \omega) - \nonumber\\
- (1 - f_{\varepsilon,
\sigma})f_{\varepsilon',-\sigma}{\cal N}_{\omega}
\delta(\varepsilon - \varepsilon' -
\hbar \omega_{\alpha}
)  ) = 0
\end{eqnarray}
Here $\nu$ is electron density of states normalized with respect
to elementary cell. One notes that in a view of the normalizing
factors of electron wave functions we have used in course of
matrix element calculation one should also put $\sum_{\bf k}
\propto t_a$, $\sum_{\bf k'} \propto t_a$ and the corresponding
factors eliminate the normalizing factor $t_a^2$ arising in course
of our estimates of the matrix element which is taken into account
in Eq.\ref{rate1}.

Note than if one would write the similar equation
(integrated over the angular variables)
in the case when momentum conservation over $x$
is allowed (when $\Delta k_F = 0$ the resulting
value of denominator in Eq.\ref{kx}
would be limited only by the
smallest possible wave vector along $x$ that is $k_x \simeq \pi/t_a$
and as a result of integration over $\xi$   one would have
$$\frac{a}{t_a^2k_F} $$
Correspondingly, an
integration over angular variables would give a factor
$$\frac{k_F}{q}$$
because in the limit $q \rightarrow 0$ in the elastic limit
one has a divergence of the corresponding cross-section.
Thus the matrix element within the equation \ref{rate1}
would have a form
\begin{equation}
\tilde {|\cal M|}^2 = J^2 \frac{k_F}{q}
\end{equation}
that is a decrease of the magnon-electron rate
with respect to a case when momentum conservation
is obeyed is descrimed by
a factor
$$ \frac{q}{(\Delta k_F)}t_a $$

One notes that the finite electron-magnon coupling strength for
low magnon frequencies (and relatively small biases) is in our
case achieved due to boundary-induced momentum non-conservation.
One expects qualitatively similar effect from any factors breaking
momentum conservation. The effect of alloy disorder significantly
increasing an efficiency of magnon-electron processes in
low-frequency region was considered in \cite{Fert}, although in
this case the disorder was related to on-site exchange constant.
We believe that the elastic scattering of electrons can also
enhance the coupling.

Let us first consider a role of this scattering when the analyzer
is still thin with respect to the elastic mean free path $l_e$.
One notes that a presence of elastic scatterers leads to a modification of
electron wave functions with respect to plane waves, due to the scattering.
Each of the scatterers produces the scattered wave, so the plane wave in the
lowest approximation with respect to the scattering processes (which
holds when $t_a < l_e$) as
\begin{equation}
e^{i{\bf kr}} \rightarrow e^{i{\bf kr}} + \sum_i f_i \frac{ \exp
{i{\bf kR_{i}} – k|{\bf R_{i} – r}|}}{|{\bf R_{i} – r}|}
\end{equation}
where $i$ numerates the scatterers, $\bf R_{i}$ is the coordinate of a scatterer
while $f$ is the scattering amplitude. One notes that the replacement of the plane waves
in Eq.\ref{initial} by the modified ones gives a non-zero result even with no account of the boundaries.
Since the magnon wave vector is assumed to be much less than the electron ones
the integrand in the Eq.\ref{initial} for each of the scatterer has a form
\begin{equation}
e^{i {\bf k r}} \cdot e^{-i({\bf k' R_i} + k'|{\bf r – R_i}|)}  =
e^{i({\bf k r'} + kr')} \cdot e^{i\Delta k r}\cdot e^{i{(\bf k –
k')R_i}}
\end{equation}
where $\bf r' = r – R_i$.
With an assumption $\Delta k << k$ the integration over $\bf r$ gives
\begin{equation}\label{part}
\frac{A}{k\Delta k}\exp i({\bf k – k'}){\bf R_i}
\end{equation}
where $|A| \sim 1$.
The total matrix element is a sum over all the scatterers in the sample.
However in the square of this element the products of contributions of
different scatterers vanish due to strongly oscillating factor in Eq.\ref{part}
depending on the scatterer coordinate. As a result, one finally obtains the following estimate:
\begin{equation}\label{scat}
|{\cal M}|^2 \sim \frac{\varepsilon_F^2}{k_Fl_e}
\end{equation}
For this estimate we have taken into account that $f^2N_i \sim
l_e^{-1}$ and that $J^2 \sim \varepsilon_F^2(k_F/\Delta k)^2$. As
it is seen, for $t_a < l_e$ the contribution of scattering is
smaller than the effect of boundaries. However one expects that
the result of Eq.\ref{scat} holds even if for the diffusive
transport in the analyzer. Indeed, one notes that actually the
contribution of each of the scatterer is formed at distances
$\sim 1/(\Delta k)$ from the scatterer. So the derivation implies
a ballistic transport only at this spatial scale which is assumed
to be less than $t_a$.

Note that in this case the electron-induced magnon relaxation rate
with an account of the fact that $\delta k \sim k_F (J/\varepsilon_F)$
can be estimated as
\begin{equation}
\tau_m^{-1} \simeq \frac{1}{k_Fl_e}\omega
\end{equation}
and thus the factor $(1/k_F l_e)$ plays a role
of Gilbert damping parameter.

\end{document}